\documentclass[twocolumn,prl,tightenlines,superscriptaddress]{revtex4-2}

\usepackage{hyperref}

\usepackage{amsmath}
\usepackage{amssymb,amsfonts,latexsym}
\usepackage{bm}
\usepackage[mathcal]{euscript}
\usepackage{graphicx}
\usepackage{epsfig}
\usepackage{color}
\usepackage{xfrac}
\usepackage{mathrsfs}

\newcommand{\rev}[1]{{\color{black} #1}}

\newcommand{\rmd}{{\rm d}}
\newcommand{\tT}{\tilde{T}}
\newcommand{\calP}{{\cal P}}

\newcommand{\Teff}{T_{\rm eff}}
\newcommand{\calQ}{{\cal Q}}
\newcommand{\calL}{{\cal L}}
\newcommand{\dtheta}{\Omega} 

\newcommand{\pathbib}{../..} 

\begin{document}

\title{XY Model with Persistent Noise} 

\author{Xia-qing Shi}
\affiliation{Center for Soft Condensed Matter Physics and Interdisciplinary Research \& School of Physical Science and Technology, Soochow University, Suzhou 215006, China}

\author{Hugues Chat\'{e}}
\affiliation{Service de Physique de l'Etat Condens\'e, CEA, CNRS Universit\'e Paris-Saclay, CEA-Saclay, 91191 Gif-sur-Yvette, France}
\affiliation{Computational Science Research Center, Beijing 100094, China}

\author{Beno\^it Mahault}
\affiliation{Max Planck Institute for Dynamics and Self-Organization (MPI-DS), 37077 G\"ottingen, Germany}
\affiliation{Laboratoire Charles Coulomb (L2C), UMR 5221 CNRS—Universit\'e de Montpellier, Montpellier F-34095, France}

\date{\today}

\begin{abstract}
We consider a 2D XY model subjected to time-correlated noise, 
a model of direct relevance to active crystals, 
which were shown recently to be able to support very large deformations without melting in the presence of persistent fluctuations.
We find that our persistent XY model can remain quasi-ordered in spite of correlations decaying much faster than allowed in equilibrium. 
We then investigate theoretically and numerically the order-disorder transition and conclude that it remains of the 
Berezinskii-Kosterlitz-Thouless type, but with scaling exponents that vary with 
the persistence time of the noise.
\end{abstract}

\maketitle

The XY model is one of the most important and most studied in equilibrium statistical physics. 
It consists of U(1) symmetric spins interacting locally via alignment.
The main features of its phase diagram in two space dimensions (2D) are well-known: 
it cannot display true long-range order but only quasi-long-range order, 
as proven by the celebrated Hohenberg-Mermin-Wagner theorem \cite{hohenberg1967existence,mermin1966absence}: 
at low enough temperature $T$, spin wave fluctuations induce a slow algebraic decay of correlations in space and time, 
with decay exponents that increase continuously with $T$.
High temperatures create a disordered phase with only short-range order and exponentially decaying correlations. 
The Berezinskii-Kosterlitz-Thouless (BKT) transition \cite{berezinskii1971destruction,kosterlitz1973ordering}, 
defined by the unbinding of the pairs of topological defects nucleated by fluctuations, separates the two regimes.

The melting of 2D crystals is one of the many topics where the XY model is relevant. 
 The celebrated Kosterlitz-Thouless-Halperin-Nelson-Young (KTHNY) theory
\cite{kosterlitz1972long,kosterlitz1973ordering,halperin1978theory,nelson1979dislocation,young1979melting} 
predicts that melting to a liquid phase, when continuous, occurs via two BKT-like transitions 
delimiting an intermediate phase with short-range positional order and quasi-long-range bond order.
KTHNY theory also predicts upper bounds for the decay exponents of positional order in the crystal phase and of 
bond order in the intermediate, usually hexatic, phase. 

In a recent publication \cite{shi2023extreme}, we showed that 2D crystals made 
of active particles subjected to pairwise repulsion forces 
can experience extremely large spontaneous deformations without melting. 
Such active crystals exhibit long-range bond order and algebraically-decaying positional order,
but with decay exponents not limited by the bounds given by KTHNY theory
\cite{kosterlitz1972long,kosterlitz1973ordering,halperin1978theory,nelson1979dislocation,young1979melting,strandburg1988two-dimensional}.
The root of these phenomena was argued to ultimately lie in the 
time-persistence of the orientation of intrinsic axes of the active particles forming the crystal.

Here we pursue this idea and consider a canonical 2D XY model in which 
 the  Gaussian white  noise acting on each spin is replaced by 
 time-persistent fluctuations governed by an Ornstein-Uhlenbeck process.
We show that, as expected from our understanding of active crystals, 
the transition separating the region of quasi-long-range order from the disordered phase
is delayed by the time-correlations of the noise, allowing to observe much stronger spin waves than in equilibrium,
without proliferation of defects. 
\rev{We then show that the order-disorder transition retrains typical characteristics of the BKT scenario, 
such as a correlation length that diverges exponentially,
while it exhibits scaling exponents
that depend on the persistence time of fluctuations.}

\begin{figure*}[t!]
   \includegraphics[width=\textwidth]{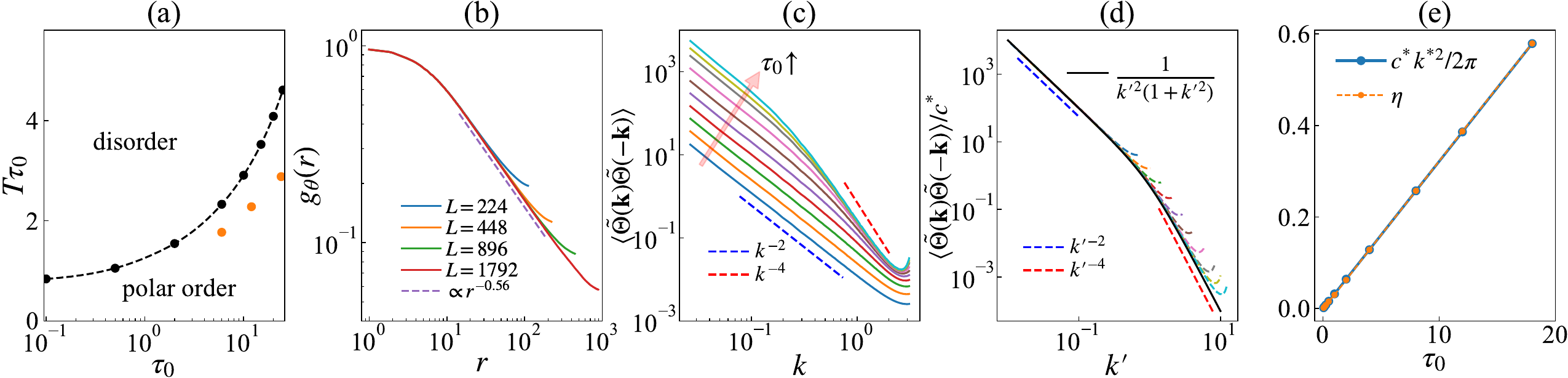}
    \caption{Quasi-ordered phase of our persistent XY model. 
     (a) phase diagram in the ${(\tau_0,T\tau_0)}$ plane. Black points and dashed line linking them: 
     order-disorder transition defined to be at inflection points of polar order parameter curves $p(T)$ in a system of size $L=448$; orange points: asymptotic BKT-like transition points determined in the second part of the paper, as summarized in  
     Table~\ref{Exponents}.
    (b) Angular correlation function $g_\theta(r)$ for $\tau_0=20$, $T=0.15$ and various system sizes.
    (c) Spatial spectra of angular field at various $\tau_0$ values ($T=0.175$). From bottom to top, $\tau_0 =0.06$, $0.12$, $0.25$, $0.5$, $1$, $2$, $4$, $8$, $12$, $18$.
    (d) Same spectra as in (c), but rescaled as explained in the main text.
    (e) Variation of  exponent $\eta$ with $\tau_0$ from direct measurement as in (b), 
    and from expression $c^* {k^*}^2/2\pi$.
    System size in (c-e) is $L=448$. Typical simulation time for each run in (b-e) is of the order of $10^7$.
 }
\label{fig1}
\end{figure*}

We consider a triangular lattice of spins $i$ whose orientation angles $\theta_i$ align locally and 
are subjected to time-correlated noise. The equation of motion for $\theta_i$ reads
\begin{align}
\dot{\theta}_i = -\partial_{\theta_i}U + \varpi_i,
\text{ with } \; \tau_0\dot{\varpi_i} = -\varpi_i +\sqrt{2\tau_0 T} \,\xi_i,  \label{XY}
\end{align}
where $U = -\tfrac{1}{2}\kappa_0 \sum_{i,j} J_{ij} \cos(\theta_i - \theta_j)$, with $\kappa_0$ the alignment elastic constant,
while $J_{ij} = 1$ if $j$ is among the 6 nearest neighbors of $i$, and $J_{ij}=0$ otherwise.
The equation governing $\varpi_i$, where $\xi_i$ is a unit-variance Gaussian white noise,  
is an Ornstein-Uhlenbeck process where $\tau_0$  is the 
persistence time of the perturbations felt by spin $i$. 
At finite $\tau_0$, $T$ can be seen as a local temperature of spinners. 
The noise amplitude on $\varpi_i$ is thus fixed by $T$, regardless of the correlation time $\tau_0$.

This out-of-equilibrium model has been considered before in \cite{paoluzzi2018effective}, 
where the focus was mostly on the perturbative effects introduced in the small $\tau_0$ limit. 
There, the equation for $\varpi$ was written $\tau_0\dot{\varpi_i} = -\varpi_i +\sqrt{2{\cal T}} \,\xi_i$, a formulation
in which the equilibrium limit $\tau_0\to 0$ is transparent, with ${\cal T}=\tau_0 T$ the thermodynamic temperature.
Here, we consider arbitrary values of $\tau_0$. 
As in equilibrium, and as shown below, a quasi-long-range ordered phase with continuously-varying
exponents is observed at low-enough $T$ values.
A rough phase diagram in the $(\tau_0,\tau_0 T)$ plane, obtained at fixed system size by 
measuring the polar order parameter 
$p=\langle m(t) \rangle_t$ where $m(t)=|\langle \exp[i\theta_j(t)] \rangle_j |$, is given in Fig.~\ref{fig1}(a).

Equation~\eqref{XY} describes active Ornstein-Uhlenbeck particles studied, e.g., in~\cite{martin2021statistical,fodor2016how}.
Following these previous approaches, we calculate the joint distribution of the phase and 
the total angular velocities $\dtheta_i = \dot{\theta}_i$ in steady-state.
Below, we sketch the main steps of the derivation, while additional details are given in Appendix D.

Taking the time derivative of $\dtheta_i$, its dynamic follows
\begin{equation} \label{eq_dtheta}
\tau_0 \dot{\dtheta}_i = -\dtheta_i - \Big[1 + \tau_0 \dtheta_j\partial_{\theta_j} \Big] \partial_{\theta_i} U + \sqrt{2 T \tau_0}\xi_i ,
\end{equation}
(hereafter summation over repeated indices is implied).
In the equilibrium limit $\tau_0 \to 0$, 
the Fokker-Planck equation governing the many-body distribution $\calP[\{\theta_i,\dtheta_i\}]$ is easily solved in steady state,
leading to the expected Boltzmann weight $\calP_{\rm eq}[\{\theta_i,\dtheta_i\}] \propto \exp[-U/(\tau_0 T) - \tfrac{1}{2}\dtheta_i^2/T]$.

For arbitrary $\tau_0 > 0$, we first analyze the quasi-ordered regime where variations of the phases are smooth and
$U \simeq U_0 + \tfrac{1}{4}\kappa_0\sum_{ij}J_{ij}(\theta_i - \theta_j)^2$. One then has
\begin{equation} \label{eq_PSW}
\calP[\{\theta_i,\dtheta_i\}] \propto \exp\left[-\frac{U}{\tau_0 T} - \frac{(\partial_{\theta_i} U)^2 + \dtheta_i \Sigma_{ij} \dtheta_j}{2T} \right],
\end{equation}
where $\Sigma_{ij} = \delta_{ij} +\kappa_0\tau_0 (6\delta_{ij} - J_{ij})$.
Since the matrix $\bm \Sigma$ is independent of the phases, integrating out the $\dtheta_i$ variables  contributes a constant prefactor.
Considering the coarse-grained field $\theta_i \to \Theta({\bf r}_i)$, such that $U = \tfrac{\kappa}{2}\int\rmd^2r |\nabla\Theta({\bf r})|^2$ 
and $(\partial_{\theta_i} U)^2 = \kappa^2\ell_0^2 \int\rmd^2r \, |\nabla^2\Theta({\bf r})|^2$
with $\kappa = \sqrt{3}\kappa_0$ and $\ell_0$ the lattice spacing,
the phase field distribution is given by
\begin{equation} \label{eq_Ptheta_SW}
P[\Theta] \propto \exp\left(-\frac{\kappa}{2\tau_0 T} \int\rmd^2 r |\nabla\Theta|^2 + \tau_0 \kappa \ell_0^2 |\nabla^2\Theta|^2 \right).
\end{equation}
Equation~\eqref{eq_Ptheta_SW} is exact at all orders in $\tau_0$~\footnote{At and near the transition, the 
renormalized value of $\kappa$ should be used. See, e.g., \cite{Kosterlitz_2016}.}.
It implies that on large scales 
the fluctuations of $\Theta({\bf r})$ are Gaussian and described by an effective temperature $\tau_0 T$.
The autocorrelation of $\Theta({\bf r})$ in Fourier space reads:
\begin{equation}
\langle\tilde{\Theta}({\bf k}) \Tilde{\Theta}({\bf -k}) \rangle = 
\frac{1}{1+\kappa \tau_0 \ell_0^2 k^2} \frac{\tau_0 T/\kappa}{k^2} \;.
\label{correl}
\end{equation} 
from which follows the angular correlation function~\cite{berthier2001nonequilibrium}
\begin{equation}
\!\! g_\theta (r)  \!=\! \langle \cos [ \Theta({\bf r}) \!-\! \Theta(0)] \rangle
 \underset{r\to\infty}{\sim} \left( \frac{r}{\ell_\tau}\right)^{-\eta}
{\rm with}\,\,
\eta\!=\!\frac{\tau_0 T}{2\pi\kappa}
\label{eta}
\end{equation}
and $\ell_\tau \sim \sqrt{\kappa \tau_0}\ell_0$ when it is much larger than $\ell_0$. 

\begin{figure*}[t!]
   \includegraphics[width=17.2cm]{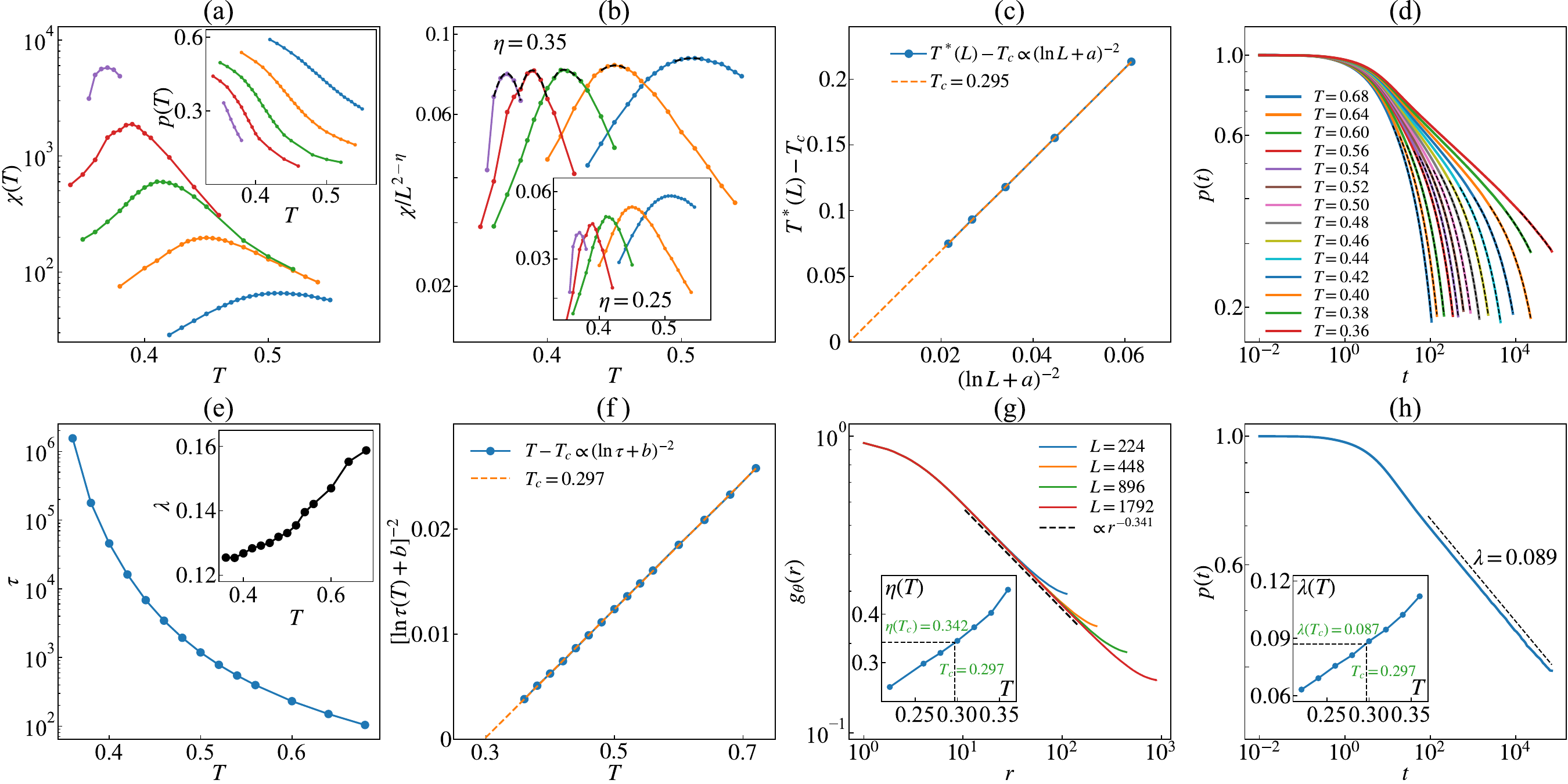}
    \caption{Order-disorder transition for $\tau_0=6$.
    (a)  $\chi(T)$ curves at different system sizes $L=56, 112, 224, 448, 896$ from blue to violet, 
    varying $T$ around the susceptibility peak. Typical simulation time for each run is of the order of $5\times10^8$. 
    (inset: corresponding order parameter curves $p(T)$).
    (b)  Same data as in (a), but rescaled by exponent $\eta=0.35$ (inset: rescaling by equilibrium value $\eta=\tfrac{1}{4}$).
    (c)  Optimal scaling of the susceptibility peak location $T^*(L)$ yielding $T_c=0.295(5)$. 
    (d)  Decay of $p(t)$ from ordered initial conditions at different $T>T_c$ values 
    (superimposed dashed lines: portions fitted by $p(t)\sim t^{-\lambda}\exp(-t/\tau)$).
    (e)  Decay time $\tau$ and exponent $\lambda$ (inset) vs $T$  from data and fits shown in (d).
    (f)  Best fit of divergence of $\tau$ near $T_c$ ($T-T_c \sim (\ln\tau+b)^{-2}$ yielding $T_c=0.297(3)$. 
    (g)  Correlation function $g_\theta(r)$ at $T=0.295$ for various system sizes, yielding $\eta=0.341(1)$ 
    (inset: $\eta$ for various $T$ values for $L=448$).
    (h) Decay of $p(t)$ for $T=0.295$  and $L=3360$ giving $\lambda\simeq 0.089(1)$
     (inset: $\lambda$ for various $T$ values).
    }
\label{fig2}
\end{figure*}

The above results are confirmed numerically. 
Simulations are performed on a triangular lattice
    containing $n\times n$ elementary blocks of $7\times 8$ sites making an almost-square 
    domain with periodic boundary conditions. 
   Below we use $L=7n$ as the linear size of the systems studied.
The model has  only two independent parameters, which we choose to be $T$ and $\tau_0$, fixing $\kappa_0=0.5$ and $\ell_0 = 1$.
   A simple Euler explicit scheme, with timestep $0.01$ is used. 
    
First of all, $\eta$,
measured directly from the decay of the correlation function $g_\theta(r)$, 
is easily observed to take values much larger than $\tfrac{1}{4}$,
the value marking the BKT transition [see one example in Fig.~\ref{fig1}(b)]. 

Spatial spectra of the angular field have the form predicted by Eq.~\eqref{correl}, developing a 
$\tau_0$-independent $1/k^4$ region at short scales while the amplitude of their large-scale $1/k^2$ 
region increases with $\tau_0$ [Fig.~\ref{fig1}(c)].
Using $k^*=\ell_0^{-1}/\sqrt{\kappa \tau_0}$ and $c^*=\tau_0^2 T$, defining $k'=k/k^*$, the angular field spectra
 normalized by $c^*$ at different $\tau_0$ values collapse nearly perfectly onto the master curve $1/[k'^2 (1+k'^2)]$
[Fig.~\ref{fig1}(d)].
Direct estimates of $\eta$ values [from data such as Fig.~\ref{fig1}(b)] and $\eta$ values calculated from
 $c^* {k^*}^2/2\pi=\tau_0 T/(2\pi\kappa)$ [the expression found in \eqref{eta}] coincide 
 and increase linearly with $\tau_0$ [Fig.~\ref{fig1}(e)].

The above results thus confirm the intuition built from our study of active crystals \cite{shi2023extreme}: 
the mere persistence in time of fluctuations extends the region of quasi-long-range order. 
The exponent $\eta$ can take large values well beyond the equilibrium $\tfrac{1}{4}$ bound. 
All this was found in the spinwave regime where defects are essentially absent and play no role. 
In equilibrium, it is well-known that in this regime, the dynamical exponent $z=2$ 
irrespective of the value of $\eta$~\cite{berthier2001nonequilibrium}.  
This is still true for our persistent XY model:
We performed quenches into the quasi-ordered phase from a perfectly ordered initial condition.
As expected, we observe, before saturation at some finite value, an algebraic decay of $p$ with an exponent
$\lambda$ and find that $\lambda\simeq\tfrac{\eta}{4}$ independently of $\tau_0$, 
i.e. $z=2$ since $z= \tfrac{\eta}{2\lambda}$ (see Appendix~A in End Matter).

We now investigate the order-disorder transition. 
The phase diagram of Fig.~\ref{fig1}(a) is of course subjected to strong finite-size effects (see below),
but gives a reasonable idea of where the transition line is located.  
Deep in the ordered phase, such as the regimes shown in Fig.~\ref{fig1}(c-e),
we see no defects at all or only rare nucleations of short lived $\pm1$ pairs.
But many defects are present near the transition and in the disordered phase, 
which induces strong fluctuations of the phases near their cores, leading to the breakdown of the spinwave approximation~\cite{pismen1999vortices}.
To understand how the persistence of the noise affects the nonlinear regime, 
we derive the phase distribution perturbatively in $\tau_0$.
As detailed in Appendix D,  $\calP$ is given at leading order by
\begin{equation*}
\calP \propto \exp\left[-\frac{U}{\tau_0 T}  
- \frac{(\partial_{\theta_i} U)^2 + \dtheta_i \Sigma_{ij} \dtheta_j - 3\tau_0T \partial_{\theta_i}^2 U}{2T} \right]~,
\end{equation*}
where the matrix $\Sigma_{ij} = \delta_{ij} +\tau_0 \partial_{\theta_i\theta_j}^2 U$ now depends on the phases,
while the last term is not constant and cannot be discarded.
Performing the Gaussian integral over the $\dtheta_i$ variables and noting that $\partial_{\theta_i}^2 U = -2 U$, we obtain 
\begin{equation}
 \label{eq_PO1_theta}
P[\{\theta_i\}] \propto \exp\left[-\frac{U}{T_{\rm eff}} - \frac{(\partial_{\theta_i} U)^2}{2 T} \right]~,
\end{equation}
where $\Teff = \tau_0 T/(1 + 2 \tau_0^2 T)$ is an effective temperature induced by the persistence of the noise.
\rev{The first term inside the exponential in Eq.~\eqref{eq_PO1_theta} dominates the dynamics of defects~\footnote{The nonlinear term $\propto (\partial_{\theta_i}U)^2 \simeq \int\rmd^2r |\nabla^2\Theta({\bf r})|^2$ should not be much sensitive to the presence of defects, since their orientation profile solves $\nabla^2\Theta = 0$~\cite{romano2023dynamical}.},
such that we expect their statistics to be equilibrium-like with an effective temperature $T_{\rm eff}$.
We thus anticipate a BKT-like transition, characterized by the relation $\Teff \simeq T_{\rm BKT} = \tfrac{1}{2}\kappa \pi$ in terms of the renormalized model parameters~\cite{kosterlitz1974critical}.}
Using the explicit expression of $\Teff$, $T_c$ and $\eta$ at the transition are found to be equal to
\begin{equation} \label{eq_Tc}
{\cal T}_c= \tau_0 T_c = \frac{\kappa \pi}{2}(1 + \kappa\pi\tau_0), \quad \eta(T_c) = \frac{1}{4}(1 + \kappa\pi\tau_0).
\end{equation}
Consistent with the phase diagram of Fig.~\ref{fig1}(a), the large-scale effective temperature $\tau_0T_c$ grows with $\tau_0$.
While the validity of Eqs.~(\ref{eq_PO1_theta},\ref{eq_Tc}) is in principle restricted to small $\tau_0$ values, 
we show below that they qualitatively capture the scenario of the transition, even for large $\tau_0$.

A direct numerical study of unbinding statistics is a very difficult task 
since there is no intrinsic and practical definition of a free defect~\footnote{Most published studies
use somewhat arbitrary thresholds to distinguish them from bound pairs (see, e.g., 
\cite{pertsinidis2001equilibrium,qi2014two-stage,digregorio2022unified}), 
something prone to producing `false positives' near and above the transition.}.
Here, we follow other routes, tested in equilibrium, to study the transition. 
We mostly focussed on the case $\tau_0=6$, but also gathered data in the equilibrium $\tau_0=0$ limit, for comparison. 

We first measured the steady-state polar order parameter $p$ and the associated 
susceptibility $\chi=L^2(\langle m^2\rangle_t - p^2)$ varying $T$ across the transition region.
Increasing system size, the $p(T)$ curves become steeper, and $\chi(T)$ curves display sharper peaks 
[Fig.~\ref{fig2}(a)], in qualitative agreement with the equilibrium case (Appendix~B, Fig.~\ref{fig3}).
In equilibrium, the peak value scales with system size as $\chi_{\rm max}(L)\sim L^{2-\eta}$ with $\eta=\tfrac{1}{4}$
\cite{kosterlitz1973ordering,kosterlitz1974critical,gupta1988phase}.
This is indeed what we obtain from our equilibrium data [Fig.~\ref{fig3}(b)], but for $\tau_0=6$ we find that 
$\eta\simeq 0.35$ [Fig.~\ref{fig2}(b)].

Our data are reliable (good statistics was cumulated to insure accurate coverage of the susceptibility peak region), 
but they remain prone to finite-size effects. 
For the accessible system sizes,
the order parameter curves and the susceptibility peaks still shift significantly as $L$ is increased. 
In the  BKT picture, the location $T^*(L)$ of $\chi_{\rm max}$ converges to the asymptotic threshold $T_c$  
like $T^*(L)-T_c \sim (\ln L + a)^{-1/\nu}$ with $\nu=\tfrac{1}{2}$ 
reflecting the square-root essential singularity of the divergence of the correlation length as $T\to T_c$ from above
\cite{dukovski2002invaded,jelic2011quench}. 
Our data follows nicely this functional form, 
which provides the estimate $T_c=0.295(5)$ 
and indicates that they are probably within the scaling region [Fig.~\ref{fig2}(c)].

We also studied quenches from the ordered phase to the transition region on the disorder side, 
another method known to provide good estimates, in equilibrium, of $T_c$ 
and of the scaling exponent $\lambda=\tfrac{\eta}{2z}$ at threshold \cite{zheng1999dynamic,ozeki2003nonequilibrium,zheng2003corrections,echinaka2016improved}.
Using large system sizes and about $\sim200-500$ samples for each $T$ value, 
one can make sure that data are clean and  free from finite-size effects until fairly large times. 
In equilibrium, for $T\gtrsim T_c$, the power law relaxation of $p$ [with exponent $\lambda(T)$]
is followed by a final exponential decay with a typical timescale $\tau(T)$ that diverges when approaching $T_c$ from above.
We observe the same behavior 
[Fig.~\ref{fig2}(d)].
For those values of $T$ furthest away from $T_c$, $\tau$ can be directly measured from the exponential decay 
of $p$ at large times. Near $T_c$, only the algebraic decay is accessible. 
Fits of our data yield accurate, system size-independent estimates of $\tau$ and rougher estimates
of $\lambda$ at various $T$ values [Fig.~\ref{fig2}(e)]. 
In equilibrium, $\tau$ is expected to scale like  $T-T_c \sim (\ln \tau + b)^{-1/\nu}$ with, again, $\nu=\tfrac{1}{2}$. 
Our data are excellently fitted by the above form, providing another estimate $T_c=0.297(3)$ 
 [Fig.~\ref{fig2}(f)]. 
  
The two values of $T_c$ obtained (0.295 and 0.297, respectively from susceptibility and quench data) 
are very close to each other, with the one derived from quench data probably more accurate.
This strengthens our confidence in them and provides a {\it de facto} error bar. 
But both were obtained assuming that the BKT exponent $\nu=\tfrac{1}{2}$ 
holds in the out of equilibrium context of our persistent XY model, something that is not guaranteed {\it a priori}.
Estimating $\nu$ directly is actually a formidable numerical task even in equilibrium, as the fits involved remain equally 
good for a range of $\nu$ values. Nevertheless, one can hope to determine the $\nu$ values providing the most coherent
answers. (See \cite{mahault2018self} for a similar approach.)
We performed the fits leading to our estimates of $T_c$
varying $\nu$ over a large range of values. As expected the fits remain very good, and the estimates of $T_c$ vary with $\nu$. 
Remarkably, the curves $T_c(\nu)$ obtained from susceptibility data and from quench data 
cross each other near $\nu=\tfrac{1}{2}$ (Appendix~C, Fig.~\ref{fig4}). 
This indicates that the BKT value $\nu=\tfrac{1}{2}$ is the one providing the most consistent interpretation of our data.

\begin{table}[t!]
 \centering
\caption{Estimates of order-disorder transition temperature ${\cal T}_c$ (${\cal T}_c = \tau_0 T_c$ for $\tau_0>0$) 
and associated exponent values
for different values of $\tau_0$. Asymptotic critical points are reported as orange dots in the phase 
diagram of Fig.~\ref{fig1}(a).
}
\begin{ruledtabular}
\begin{tabular}{lcccc}
 & ${\cal T}_c$ & $\eta$ & $\lambda$ & $z$ \\
  $\tau_0=0$ (equilibrium) & 0.727(2) & 0.232(2) &  0.059(2) & 1.97(3) \\
  $\tau_0=6$                       & 1.78(2) & 0.342(4) & 0.087(2) &  1.97(3) \\
  $\tau_0=12$                     & 2.28(3) & 0.43(1) & 0.110(2) &  1.98(3)\\ 
  $\tau_0=24$                     & 2.88(7) & 0.54(1) & 0.137(3) &  1.97(3)
  \end{tabular}
 \end{ruledtabular}
 \label{Exponents}
\end{table}

Being confident that $T_c\in [0.295,0.300]$ for our persistent XY model with $\tau_0=6$, 
we can now provide estimates of the scaling exponents $\eta$ and $\lambda$, 
and thus of $z=\tfrac{\eta}{2\lambda}$ at the transition [Fig.~\ref{fig2}(g,h)]. 
The decay of $g_\theta(r)$ at $T_c$ yields $\eta=0.342(4)$, 
in agreement with the estimate obtained from the scaling of the susceptibility peak.
The decay of $p(t)$ at $T_c$ yields $\lambda=0.087(2)$. It follows that we estimate $z=1.97(3)$. 

We applied the same methodology to estimate these exponents in the equilibrium limit $\tau_0=0$
(Appendix~B, Fig.~\ref{fig3}). 
This yielded ${\cal T}_c=0.727(2)$, $\eta=0.232(2)$, and $\lambda=0.059(2)$. 
We thus estimate $z=1.97(3)$ (or $z=2.11(7)$ if the BKT value $\eta=\tfrac{1}{4}$ is used). 
These estimates are less accurate than those obtained for $\tau_0=6$, for which a much larger
numerical effort was provided, they are close the best or exact equilibrium BKT values ($\eta=0.25$, $z\simeq 2$, 
see \cite{dukovski2002invaded,jensen2000dynamic}).
Our key results are summarized in Table~\ref{Exponents}, which also includes preliminary estimates at $\tau_0=12$ and $24$ obtained from quench experiments.

To summarize, we have first shown that strong spinwaves leading to a decay of orientational order much faster than
allowed in equilibrium can be observed even deep in the quasi-ordered phase of our persistent XY model, 
where essentially no defect is present. 

\rev{We then studied the order-disorder transition numerically. The most coherent interpretation of our results is that
the transition remains qualitatively of the BKT type, with the exponents $\nu=\tfrac{1}{2}$ and $z=2$
unchanged out of equilibrium. 
On the other hand, they unambiguously show that the exponents $\eta$ and $\lambda$ vary with $\tau_0$. 
In particular, $\tau_0 T_c$ and $\eta(T_c)$ both increase with $\tau_0$.
While this behavior is qualitatively captured by our perturbative approach (Eq.~\eqref{eq_Tc}),
a quantitative characterization of the transition would require to determine the renormalized value of 
$\kappa\tau_0$ at the BKT-like fixed point.}
Further work extending the renormalization group analysis of the BKT transition is thus needed 
to fully grasp the effects of time-persistence.

\rev{Our results are in line with our earlier findings on active crystals~\cite{shi2023extreme},
and notably imply that a very broad class of out-of-equilibrium dynamics naturally sustain defect-free, 
extreme deformations not allowed in equilibrium.
A practical consequence of this observation is that, out of equilibrium, 
the bound $\eta = \tfrac{1}{4}$ from the BKT theory cannot be used to locate the defect-unbinding transition.
Coming back to active crystals, or passive crystals subjected to an active bath~\cite{shi2024effect,Massana-Cid2024multiple}, 
our results suggest that their melting transition, when continuous, could remain qualitatively 
---but not quantitatively--- in the framework of the KTHNY theory~\cite{digregorio2022unified}. 
We hope to make progress about this difficult question in the future.}

\acknowledgments
We thank Yu Duan, Alexandre Solon, and Yongfeng Zhao for useful remarks and a careful reading of the manuscript.
This work is supported by the National Natural Science Foundation of China (Grants No. 12275188 and No. 11922506)

\bibliographystyle{apsrev4-2}
\bibliography{\pathbib/Biblio-current.bib}

\onecolumngrid

\vspace{12pt}
\noindent\hrulefill \hspace{24pt} {\bf End Matter} \hspace{24pt} \hrulefill
\vspace{12pt}

\twocolumngrid

\begin{figure}[t]
 \includegraphics[width=\columnwidth]{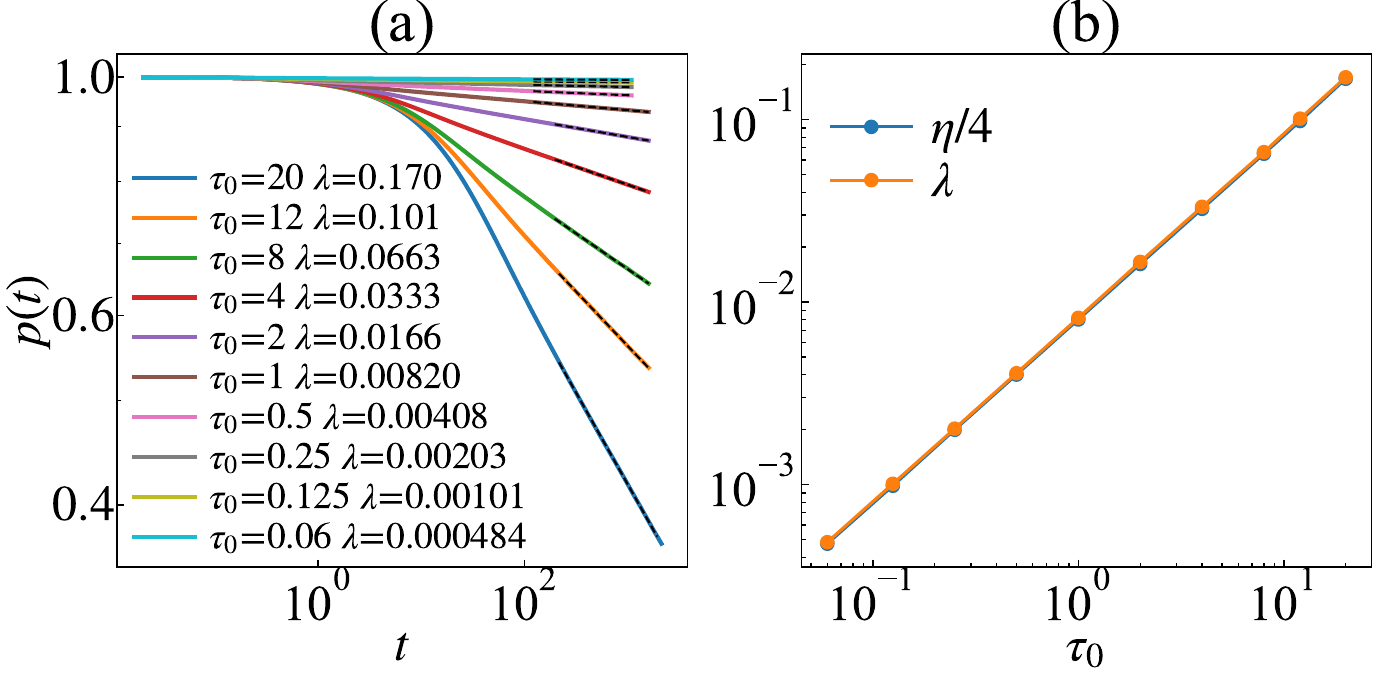}
    \caption{Quenches into the quasi-ordered phase.
    (a) Time-decay of $p$ for various $\tau_0$ values ($T=0.175$, $L=3360$); superimposed dashed lines are 
    powerlaw fits giving the estimates of $\lambda$ indicated in the legends.
    (b) Variation with $\tau_0$ of $\lambda$ [data from (a)] and $\eta/4$ [from decay of $g_\theta(r)$]. 
     }
\label{figEM1}
\end{figure}

\begin{figure*}[t]
   \includegraphics[width=\textwidth]{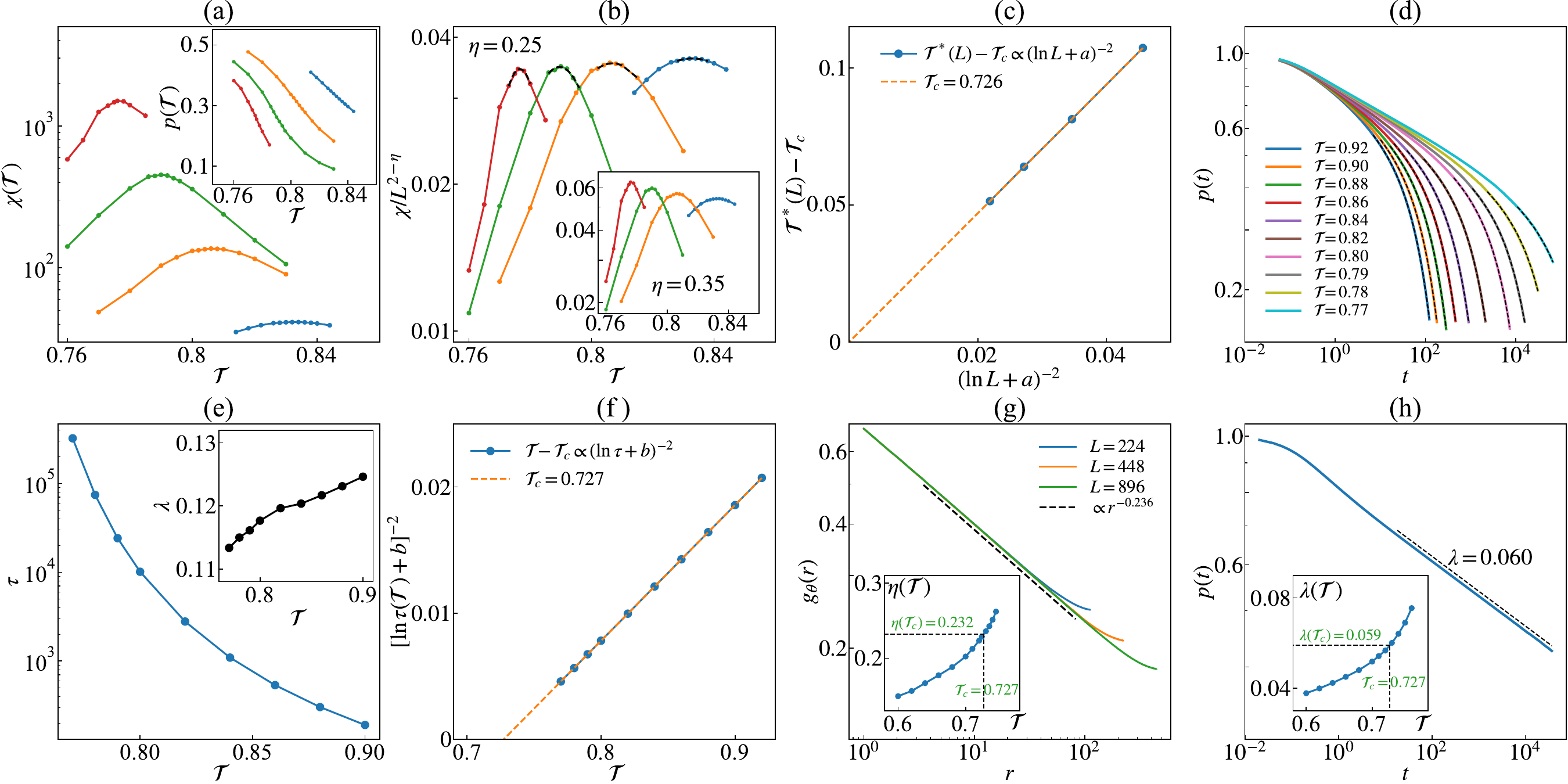}
    \caption{
    BKT transition in the equilibrium limit $\tau_0=0$, varying ${\cal T}$. 
    (a)  $\chi({\cal T})$ curves at different system sizes $L=56, 112, 224, 448$ from blue to red, 
    varying ${\cal T}$ around the susceptibility peak 
    (inset: corresponding order parameter curves $p({\cal T})$). Typical simulation time for each run is $5\times10^8$ with $dt=0.02$.
    (b)  Same data as in (a), but rescaled by exponent $\eta=0.25$ 
    (inset: rescaling by the value $\eta=0.35$ found for $\tau_0=6$).
    (c)  Optimal scaling of the susceptibility peak location ${\cal T}^*(L)$ yielding ${\cal T}_c=0.726(5)$. 
    (d)  Decay of $p(t)$ from ordered initial conditions at different ${\cal T}>{\cal T}_c$ values 
    (superimposed dashed lines: portions fitted by $p(t)\sim t^{-\lambda}\exp(-t/\tau)$).
    (e)  Decay time $\tau$ and exponent $\lambda$ (inset) vs ${\cal T}$  from data and fits shown in (d).
    (f)  Best fit of divergence of $\tau$ near ${\cal T}_c$ (${\cal T}-{\cal T}_c \sim (\ln\tau+b)^{-2}$ yielding ${\cal T}_c=0.727(2)$. 
    (g)  Correlation function $g_\theta(r)$ at ${\cal T}=0.73$ for various system sizes, yielding $\eta=0.236(1)$ 
    (inset: $\eta$ for various ${\cal T}$ values for $L=448$).
    (h) Decay of $p(t)$ for ${\cal T}=0.73$  and $L=3360$ giving $\lambda\simeq 0.060(1)$
     (inset: $\lambda$ for various ${\cal T}$ values).}
\label{fig3}
\end{figure*}

{\it Appendix A: Quenches into the quasi-ordered phase---}
We followed the evolution of $p$ starting from a perfectly aligned initial configuration with the global order oriented along one lattice axis. 
(When $\tau_0\ne 0$ we first `relax' the noise field by running the $\varpi$  equation in~\eqref{XY} for $10\,\tau_0$).
When quenching into the quasi-ordered phase, $p$ first decreases algebraically with an exponent $\lambda$, 
then saturates at some finite value (as expected in a finite-size system). 
To obtain clean-enough results, we use large system sizes and repeat the process over $\sim200-500$ samples.
Using the parameter values of Fig.~\ref{fig1}(e), 
we estimate  $\lambda$ for various $\tau_0$ values, and find these estimates undistinguishable from those of 
$\eta/4$ measured from the decay of $g_\theta(r)$ calculated in the steady-state (Fig.~\ref{figEM1}). 
This shows that $z\simeq 2$ irrespective of the value of $\tau_0$.\\

{\it Appendix B: Numerical data obtained in the equilibrium limit---}We simulated our XY model for $\tau_0=0$, using the same methodology as for the non-equilibrium case $\tau_0=6$. 
The results  are presented in Fig.~\ref{fig3}, which is arranged like Fig.~\ref{fig2}. 
Estimates of $T_c$, $\eta$, $\lambda$, and thus of $z$, are reported in Table~\ref{Exponents}.
They are in line with those reported elsewhere, and in agreement with BKT theory.\\

{\it Appendix C: Varying the BKT exponent $\nu$---}In Figs.~\ref{fig2} and \ref{fig3}, estimates of $T_c$ were obtained from the ansatz   
$T^*(L)-T_c \sim (\ln L + a)^{-1/\nu}$ for susceptibility peak data, and
$T-T_c \sim (\ln \tau + b)^{-1/\nu}$) for quench data, using the BKT value $\nu=\tfrac{1}{2}$.
In the absence of theoretical results, one has to consider that $\nu$ may take a different value.
In Fig.~\ref{fig4}(a), we show the results of the fits for various values of $\nu\in [0.2,1.2]$. 
Except  for the extreme values in this interval, the fits remain very good. 
The two estimates of $T_c$ are closest to each other near $\nu=\tfrac{1}{2}$ both out of and in equilibrium.
Moreover, the prefactors $a$ and $b$, which can be interpreted as
(the logarithms of) elementary length scales and timescales, take `reasonable' values (of order $~1$) for $\nu=\tfrac{1}{2}$
 [Fig.~\ref{fig4}(b)].\\

\begin{figure}[h]
 \includegraphics[width=\columnwidth]{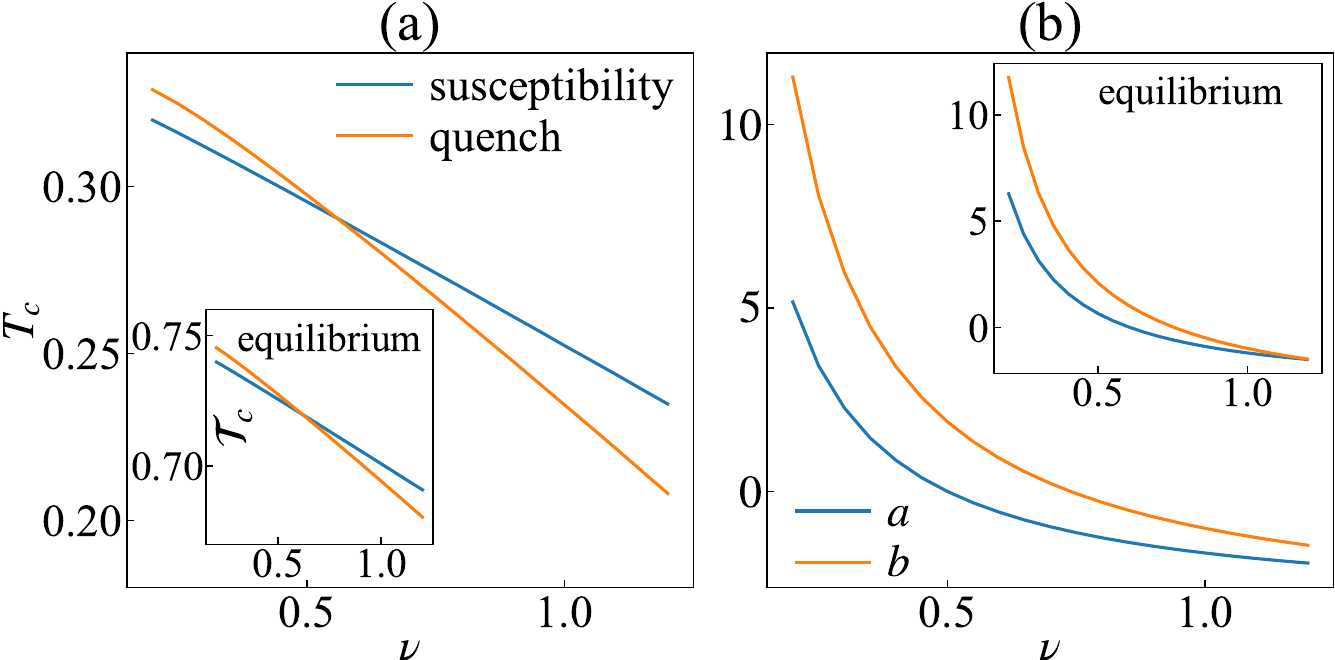}
    \caption{Varying the value of the $\nu$ exponent used in fits to estimate $T_c$ for our $\tau_0=6$ data. 
    ($T^*(L)-T_c \sim (\ln L + a)^{-1/\nu}$ for susceptibility peak data, and
    $T-T_c \sim (\ln \tau + b)^{-1/\nu}$) for quench data).
    (a) Variation of estimated $T_c$ with $\nu$.     
    (b) Variation of the prefactors $a$ and $b$ with $\nu$.
    Insets: same but for our equilibrium ($\tau_0=0$) data.
     }
\label{fig4}
\end{figure}

\setcounter{equation}{0}
\renewcommand{\theequation}{D\arabic{equation}}
{\it Appendix D: Derivation of the stationary phase distribution---}From Eq.~\eqref{eq_dtheta} of the main text, the distribution $\calP[\{\theta_i,\dtheta_i\},t]$ is governed by 
the Fokker-Planck equation $\partial_t \calP + \dtheta_i\partial_{\theta_i}\calP - \calL_{\dtheta} \calP = 0$, 
where
\begin{equation} \label{eq_FP}  
\calL_{\dtheta} = \frac{1}{\tau_0}\partial_{\dtheta_i}\left[\dtheta_i 
+ (1 + \tau_0\dtheta_j \partial_{\theta_j})\partial_{\theta_i}U\right]
+ \frac{T}{\tau_0} \partial^2_{\dtheta_i} .
\end{equation}
In steady-state, we set $\partial_t\calP = 0$, and seek solutions of the form
\begin{equation} \label{eq_ansatzP}
\calP[\{\theta_i,\dtheta_i\}] \propto \exp\left[-\frac{U}{\tau_0 T} - \frac{\dtheta_i^2}{2T} + \calQ[\{\theta_i,\dtheta_i\}]\right]~.
\end{equation}
For $\tau_0 \to 0$
it is straightforward to verify that~\eqref{eq_ansatzP} solves $\dtheta_i\partial_{\theta_i}\calP - \calL_{\dtheta} \calP = 0$ with $\calQ = 0$.
Following 
\cite{martin2021statistical}, we consider the rescaled variables $\tilde\dtheta_i = \tau_0^{1/2}\dtheta_i$ and define the effective temperature $\tT = \tau_0 T$.
To simplify the search for solutions, we further define
\begin{equation} \label{eq_ansatzQ}
\calQ = \tilde\calQ - \frac{\tau_0}{2\tT} \left[ \tilde\dtheta_i\tilde\dtheta_k \partial_{\theta_i\theta_k}^2 U + (\partial_{\theta_i} U)^2 \right]~.
\end{equation}
Replacing the ansatz~\eqref{eq_ansatzP} with~\eqref{eq_ansatzQ} into $\dtheta_i\partial_{\theta_i}\calP - \calL_{\dtheta} \calP = 0$, 
we find that for $\tau_0 > 0$ $\tilde\calQ$ satisfies 
\begin{align} 
\tT \partial^2_{\tilde\dtheta_i}\tilde\calQ & + \partial_{\tilde\dtheta_i}\tilde\calQ\left[ \tT \partial_{\tilde\dtheta_i}\tilde\calQ - \tilde\dtheta_i 
- \tau_0 \tilde\dtheta_k\partial_{\theta_i\theta_k}^2 U + \tau_0^{1/2}\partial_{\theta_i}U\right] \nonumber \\
\label{eq_tQ}
& -\tau_0^{1/2} \tilde\dtheta_i \partial_{\theta_i}\tilde\calQ + \frac{\tau_0^{3/2}}{2\tT} ({\tilde\dtheta_i}\partial_{\theta_i})^3 U = 0~.
\end{align}

In the spin wave approximation, we write the potential as $U \simeq U_0 + \tfrac{1}{4}\kappa_0\sum_{ij}J_{ij}(\theta_i - \theta_j)^2$.
Then the term of third order in the derivatives of $U$ in Eq.~\eqref{eq_tQ}, the only one not proportional to $\tilde\calQ$, vanishes, so that $\tilde\calQ = 0$
is a trivial solution of \eqref{eq_tQ}. Using (\ref{eq_ansatzP},\ref{eq_ansatzQ})
and going back to non-rescaled variables, Eq.~\eqref{eq_PSW} of the main text is recovered.

In and near the disordered phase populated by topological defects, 
angular fluctuations between neighboring sites can be strong, and the harmonic approximation of $U$ breaks down.
Then, $\partial_\theta^3 U \ne 0$ in Eq.~\eqref{eq_tQ}, which admits nontrivial solutions.
No general solution for $\tilde\calQ$ is known, and even the existence of an analytic form is not guaranteed. 
Therefore, we instead work perturbatively in $\tau_0$, and write
$\tilde\calQ[\{\theta_i,\tilde\dtheta_i\}] = \sum_{n=1}^\infty \tau_0^{n/2} \calQ_n[\{\theta_i,\tilde\dtheta_i\}]$.
Plugging this decomposition into \eqref{eq_tQ}, and matching terms of same order in $\tau_0$, 
one easily finds that $\calQ_1 = 0$,
such that the leading order contribution 
is $\calQ_2 \tau_0$.
The equation for $\calQ_2$ takes the simple form 
$\tT \partial^2_{\tilde\dtheta_i}\calQ_2 - {\tilde\dtheta_i}\partial_{\tilde\dtheta_i}\calQ_2 = 0$,
such that $\calQ_2$ is only a function of the phases.
At the next order, we obtain
\begin{equation*}
\tT \partial^2_{\tilde\dtheta_i}\calQ_3 - \tilde\dtheta_i\partial_{\tilde\dtheta_i}\calQ_3 - \tilde\dtheta_i\partial_{\theta_i}\calQ_2 + \frac{1}{2\tT}(\tilde\dtheta_i\partial_{\theta_i})^3U  = 0.
\end{equation*} 
Looking for polynomials in the $\tilde\dtheta_i$'s, we find
$\calQ_3 = (\tilde\dtheta_k\partial_{\theta_k})[ \tfrac{1}{6\tT}(\tilde\dtheta_i\partial_{\theta_i})^2U + \partial_{\theta_i}^2 U - \calQ_2]$.
At the next order, which allows to fix $\calQ_2$, we have
\begin{align*}
& \tT \partial^2_{\tilde\dtheta_i}\calQ_4 - \tilde\dtheta_i\partial_{\tilde\dtheta_i}\calQ_4
- \left[(\tilde\dtheta_k\partial_{\theta_k})^2 - (\partial_{\theta_i}U)\partial_{\theta_i}\right] \left( \partial_{\theta_k}^2U - \calQ_2 \right)\\
& - \frac{1}{6\tT}\left[(\tilde\dtheta_k\partial_{\theta_k})^2 - 3(\partial_{\theta_i}U)\partial_{\theta_i}\right](\tilde\dtheta_k\partial_{\theta_k})^2U = 0~.
\end{align*}
Solutions of this equation take the form
\begin{align*}
\calQ_4 & = -\frac{1}{4!\tT}(\tilde\dtheta_k\partial_{\theta_k})^4 U + \frac{1}{4\tT}(\partial_{\theta_i}U)\partial_{\theta_i}(\tilde\dtheta_k\partial_{\theta_k})^2U \\
& + \frac{1}{2}(\tilde\dtheta_k\partial_{\theta_k})^2\left( \calQ_2 - \frac{3}{2}\partial_{\theta_i}^2U \right) + \calQ_{4,0}~,
\end{align*}
where $\calQ_{4,0}$ is a function of the phases only.
The above expression is a solution only when the equality 
$[(\partial_{\theta_i}U)\partial_{\theta_i} - \tT\partial_{\theta_i}^2 ]\left[ \frac{3}{2}\partial_{\theta_k}^2U -\calQ_2 \right] = 0$
is verified.
We thus conclude that 
$\calQ_2 = \frac{3}{2}\partial_{\theta_i}^2U$.
Similarly, 
$\calQ_{4,0}$ is determined at order $\tau_0^3$ from the equation of $\calQ_6$. 
In fact, all terms of the series can be calculated iteratively, albeit
the corresponding expressions become increasingly complex.

At leading order $\tau_0$ the steady-state distribution reads
\begin{equation*}
\calP \propto \exp\left[-\frac{U}{\tau_0 T} 
- \frac{(\partial_{\theta_i} U)^2 + \dtheta_i \Sigma_{ij} \dtheta_j - 3\tau_0T \partial_{\theta_i}^2 U}{2T} \right]
\end{equation*}
where the matrix $\Sigma_{ij} = \delta_{ij} +\tau_0 \partial_{\theta_i\theta_j}^2 U$.
The phase distribution is then obtained by performing the Gaussian integral over the angular velocities, yielding a $\sqrt{{\rm det}\bm \Sigma^{-1}}$ prefactor.
Up to ${\cal O}(\tau_0^2)$ terms, we have
$\Sigma_{ij}^{-1} \simeq \delta_{ij} - \tau_0 \partial_{\theta_i\theta_j}^2 U \simeq e^{-\tau_0 \partial_{\theta_i\theta_j}^2 U}$
which, together with the relation ${\rm det}(e^{\bm M}) = e^{ {\rm Tr}(\bm M)}$, 
give $\sqrt{{\rm det}\bm \Sigma^{-1}} \simeq \exp[-\tfrac{\tau_0}{2}\partial_{\theta_i}^2 U]$.
Evaluating the Laplacian of the potential explicitly, we have $\partial_{\theta_i}^2 U = -2 U$,
yielding finally the expression given in Eq.~\eqref{eq_PO1_theta} of the main text.

\end{document}